%% file: 2020-VPT-EPTCS.tex
\documentclass[submission,copyright,creativecommons]{eptcs}
\providecommand{\event}{          
	Verification and Program Transformation (VPT 2020)
}
\usepackage{breakurl}             
\usepackage[strings]{underscore}  
\usepackage{color}

\newcommand{\texttts}[1]{\texttt{#1}}

\usepackage{xspace}
\usepackage{enumitem}

\usepackage{listings} 

\lstdefinestyle{myscala}{
	language=scala,     
	stringstyle=\ttfamily,
	showstringspaces = false,
	basicstyle=\linespread{0.9}\small\ttfamily,
	commentstyle=\small\emph,
	keywordstyle=\color{blue}\bfseries,
	mathescape=true,
	breaklines=true,
	xleftmargin=0em, 
	,morekeywords={ensuring,require}
	,otherkeywords={}
	,deletekeywords={true,false},
	columns=[l]flexible
}

\lstdefinestyle{myprolog}{
	,style=myscala,
	,language=prolog
	,deletekeywords={false,true}
}

\title{Transformational Verification of Quicksort}

\author{
	Emanuele De Angelis
	\institute{
		CNR-IASI\\[2pt]
		Via dei Taurini 19, 00185 Roma, Italy
	}
	\email{emanuele.deangelis@iasi.cnr.it}
	\and
	Fabio Fioravanti
	\institute{DEC, University ``G.~d'Annunzio" of Chieti-Pescara\\[2pt]
		Viale Pindaro 42, 65127 Pescara, Italy
	}
	\email{fabio.fioravanti@unich.it}
	\and
	\and
	Maurizio Proietti
	\institute{
		CNR-IASI\\[2pt]
		Via dei Taurini 19, 00185 Roma, Italy
	}
	\email{maurizio.proietti@iasi.cnr.it}
}

\def\titlerunning{Transformational Verification of Quicksort}

\def\authorrunning{E.~De Angelis, F.~Fioravanti \& M.~Proietti}

\begin{document}
	\maketitle

	\begin{abstract}
		\input{sections/0_abstr.tex}

	\end{abstract}
	
	\section{From Program Transformation to Program Verification}
	\input{sections/1_intro.tex}

	\section{Program Verification via Constrained Horn Clause Transformation}
	\input{sections/2_chcsat.tex}

	\section{Specification of Quicksort with Parameterized Catamorphisms}
	\input{sections/3_specification.tex}
	
	\section{Removing List Arguments}
	\input{sections/4_transformation.tex}

	\section{Related Work and Conclusions}
	\input{sections/5_relConcl.tex}

\input{sections/6_Biblio.bbl}

	\section*{Appendix}
	\input{sections/7_appendix.tex}
\end{document}

%% file: sections/0_abstr.tex
\begin{abstract}
Many transformation techniques developed for constraint logic programs, also known as
{\em constrained Horn clauses} (CHCs),
have found new useful applications in the field of program verification.
In this paper, we work out a nontrivial case study through the transformation-based 
verification approach. We consider the familiar \textit{Quicksort}
program for sorting lists, written in a functional programming language,
and we verify the pre/postconditions
that specify the intended correctness properties of the functions defined in the
program.
We verify these properties by: 
(1)~translating them into CHCs, 
(2)~transforming the CHCs by removing all list occurrences, and
(3)~checking the satisfiability of the transformed CHCs by using the Eldarica solver
over booleans and integers.
The transformation mentioned at Point~(2) requires an extension of the algorithms
for the elimination of inductively defined data structures presented in
previous work, because during one stage of the transformation
we use as lemmas some properties that have been proved at previous stages.
\end{abstract}

%% file: sections/1_intro.tex
{\em Program transformation} 
gained a lot of popularity after the seminal paper by Burstall and Darlington~\cite{BuD77},
who advocated an approach based on {\em transformation rules}, which preserve
the semantics of programs, and {\em transformation strategies}, which guide
the application of the rules towards a goal of interest.
This approach enables the separation, during program development, of the correctness issue from 
the efficiency issue.

Burstall and Darlington's rule-based approach has been proposed in 
the context of functional programming,
and later extended to other programming paradigms, such as
logic programming~\cite{PeP89b,TaS84} and 
constraint logic programming (CLP)~\cite{EtG96}.
The interest of applying program transformation techniques to
declarative programming languages, like functional
and logic programming, is due to the fact that in that context both specifications 
and programs are written as logical formulas, 
and program transformation can be viewed as
a means for deriving, via logical deduction, efficient programs that 
are correct by construction~\cite{Hog81}.

Starting from the late 1990s, many program analysis and transformation techniques for 
logic and constraint logic programs
have found new applications in the field of \emph{program verification}.
Initially, they have been applied to the proof of properties
for abstract computational models such as {\em Petri nets},
{\em timed automata}, and {\em infinite state transition systems}~\cite{BaG08,DeP99,Fi&01a,FrO97a,LeL00a,Ra&97},
and, later on, also for verifying programs written in concrete programming languages, 
including imperative and object-oriented languages~\cite{Al&07,De&14c,Me&07,Pe&98}.
Indeed, logic programming, possibly extended with constraint theories,
is a very suitable language for specifying program semantics and 
program properties~\cite{Gr&12,Pe&98,PoR07}.
Moreover, the notions of least and greatest models are the logical 
counterparts of least and greatest fixed points often used for program verification. 

In the field of program verification, constraint logic programs
are often called \emph{constrained Horn clauses} (CHCs), when
we want to stress their use as a reasoning formalism, rather than
as a programming language~\cite{Bj&15}.
The underlying constraint theories used in CHCs are typically those that
axiomatize data structures used in programming, such as
booleans, integer numbers, real numbers, bit vectors, arrays, heaps,
and inductively defined data structures such as lists and trees.
For checking the satisfiability of CHCs, effective \emph{solvers}, 
such as {\em Eldarica}~\cite{HoR18} and {\em Z3}~\cite{DeB08} with the 
{\em Spacer} Horn engine~\cite{Ko&13},
have been developed during the last years.

Several CHC transformations, including {\em fold/unfold} transformations and
{\em CHC-specialisation},
have been applied to verification problems~\cite{De&14c,De&17b,De&18a,KafleG17b,MoF17}.
The basic idea is to transform a set of clauses~$P$, whose satisfiability
guarantees a certain program property, into a new set of clauses $P'$, such that
the satisfiability of~$P'$ (1)~implies the satisfiability of $P$, and
(2)~is more effectively checked by the available CHC solver.
One of these CHC transformations is the fold/unfold strategy for the
elimination of inductively defined data structures
from CHCs. This strategy was first proposed as a means 
for improving the efficiency of logic programs
by avoiding intermediate data structures~\cite{PrP95a}, and is
strongly related to the well-known \emph{deforestation} transformation in 
functional programming~\cite{Wad90}.
In the context of CHC verification,
the advantage of eliminating inductively defined data structures is that the 
satisfiability of the derived clauses can be proved in simpler domains, such as 
the theory of booleans or the theory of linear arithmetic, 
for which existing solvers are very effective. 

In previous work~\cite{De&18a,De&20a}, we have shown that, by eliminating 
inductively defined data structures from CHCs, we can avoid to extend
solvers with induction-based inference rules, and yet
we can prove universally quantified properties of
programs acting on those structures.
Indeed, experiments show that our two-step technique, consisting in 
preprocessing CHCs by eliminating inductively defined data structures, and then applying
CHC solvers over booleans and integers, is competitive with respect to 
approaches based on extending solvers with induction over data structures~\cite{ReK15,Un&17}. 

In this paper, we work out a case study through the transformation-based 
verification approach. We consider a program \textit{Quicksort} for sorting lists,
written in the pure functional fragment of {\em Scala}~\cite{OderskySV11},
implementing the familiar algorithm invented by Tony Hoare~\cite{Hoare61}.
The program is equipped with {\em contracts}, i.e., pre/postconditions
that specify the intended correctness properties of the various program functions.
We check that the program verifies all contracts by: 
(1)~translating them into CHCs, 
(2)~transforming the CHCs by removing all list occurrences, and
(3)~checking the satisfiability of the transformed CHCs by using the Eldarica solver
over booleans and integers.
The transformation mentioned at Point (2) requires an extension of the algorithms
for the elimination of inductively defined data structures presented in
previous work, because during one stage of the transformation
we will use as lemmas {some} contracts that we have verified at previous stages.

The advantage of our approach is that we avoid the use of very complex
program verifiers, such as the {\sc Stainless} tool developed for Scala~\cite{HamzaVK19},
which combine reasoning in Hoare logic with induction and constraint solving,
and instead, by our transformation, we reduce the verification task 
to a problem that can be handled by simpler CHC solvers.
In fact, our specific \textit{Quicksort} verification problem is not solved by
{\sc Stainless}.

\medskip

The paper is organized as follows.
In Section~\ref{sec:partition}, we recall the transformation-based verification approach
by considering the \texttt{partition} function, which is used by the 
\textit{Quicksort} program. 
In Sections~\ref{spec} and~\ref{listrem}, we apply the transformation-based verification approach
to the whole \textit{Quicksort} program.
In particular, in Section~\ref{spec} we show how the problem of verifying the 
correctness of \textit{Quicksort} with respect to its contracts
is translated to CHCs.
Then, in Section~\ref{listrem}, we show how those CHCs
are transformed by removing all
list terms, hence deriving a set of clauses over booleans and integers
whose satisfiability is proved by Eldarica.
Finally, in Section~\ref{relwork}, we compare our contribution to
related work and we make some concluding considerations.

%% file: sections/2_chcsat.tex
\label{sec:partition}

\newcommand{\ispart}{\textit{Pivot}\xspace}

In this section, we recall the transformation-based approach to program verification
by means of a simple example.
We consider a function \texttt{partition} 
for partitioning a list of natural numbers into two sublists
by using a pivot element.
This function will be used in the \textit{Quicksort} program of Section~\ref{spec}.
We translate the \texttt{partition} function 
into a set \textit{PartitionCHCs} of clauses,
and the contract associated with \texttt{partition}
into a set \textit{Gs} of \textit{goals}, that is, clauses with \texttt{false} head.
The satisfiability of $\textit{PartitionCHCs}\cup\{\textit{G}\}$, for all $G$ in \textit{Gs},
 guarantees that \texttt{partition} is 
correct with respect to the specified contract.
Then, for all $G$ in \textit{Gs}, we apply the transformation technique based on the 
Elimination Algorithm~\cite{De&18a}
for removing list occurrences from $\textit{PartitionCHCs}\cup\{\textit{G}\}$.
The result of the transformation is a set $T_G$ of clauses
over the theories \textit{Bool} of boolean values and \textit{LIA}
of linear integer arithmetic, which is satisfiable if and only if $\textit{PartitionCHCs}\cup\{\textit{G}\}$
is satisfiable.
Finally, we check the satisfiability  of $T_G$ by using a CHC solver over \textit{Bool} and \textit{LIA}.

\smallskip
Let us consider the following program  \textit{Partition}
written in the pure functional fragment of Scala~\cite{OderskySV11}:

\begin{lstlisting}[style=myscala
%%CAPTION
 ,caption={Program \textit{Partition}. Variable \texttt{res} denotes the pair returned by the \texttt{partition} function, and
\texttt{res._1} and \texttt{res._2} denote its first and second components, respectively.}
 ,captionpos=b
 ,label={lst:partition}
 ,abovecaptionskip=6pt
 ,belowcaptionskip=4pt
 ,escapechar=|
] 
def all_grt(x: Nat, l: List[Nat]): Boolean = {
  l match {
    case Nil() => true
    case Cons(y, ys) if (x =< y) => false
    case Cons(y, ys) if (x > y) => all_grt(x, ys)
 }

def all_leq(x: Nat, l: List[Nat]): Boolean = {
  l match {
    case Nil() => true
    case Cons(y, ys) if (x > y) => false
    case Cons(y, ys) if (x =< y) => all_leq(x, ys)
  }
}

def partition(x: Nat, l: List[Nat]): (List[Nat], List[Nat]) = {
  l match {
    case Nil() => (Nil[Nat](), Nil[Nat]())
    case Cons(y, ys) =>
      val (l1, l2) = partition(x, ys)
      if (x > y) { (Cons(y, l1), l2) }
      else        { (l1, Cons(y, l2)) }
  }
} ensuring { res =>
  all_grt(x, res._1) && all_leq(x, res._2)       // partition postcondition
}
\end{lstlisting}

\noindent
Given a natural number \texttt{x} and a list \texttt{l} of natural numbers, we have that
(i) \texttt{all_grt(x,l)} (and, respectively,  \texttt{all_leq(x,l)}) returns \texttt{true} if
\texttt{x} is greater than (respectively, less than or equal to) every element of \texttt{l}, 
and \texttt{false} otherwise,
(ii) \texttt{partition(x,l)} returns a pair of lists \texttt{(l1,l2)},
where \texttt{l1} (respectively,  \texttt{l2})  is the list of all the elements \texttt{y} of \texttt{l}
such that \texttt{x} is greater than (respectively, less than or equal to) \texttt{y}.
The \texttt{partition} function 
is annotated with a postcondition, specified by the \texttt{\color{blue}\bfseries ensuring} assertion,
which encodes the following contract:  

\smallskip
\noindent
{\small 
$\forall$\texttt{x,l,l1,l2. partition(x,l)==(l1,l2) 
==> all_grt(x,l1)  \&\&   all_leq(x,l2) }}~(Contract \ispart)

\smallskip
\noindent
In general, a contract consists of a precondition, specified by a \texttt{\color{blue}\bfseries require}
assertion, and a postcondition, specified by an \texttt{\color{blue}\bfseries ensuring} assertion.
However, in the case of \texttt{partition}, the precondition is missing (i.e., it is \texttt{true}).

In order to prove that the contract \ispart is indeed satisfied, 
we first consider the translation of the \textit{Partition} program
into the following set \textit{PartitionCHCs} of clauses 
(where natural numbers have been translated into non-negative integers in the {\textit{LIA} theory}):

\begin{lstlisting}[style=myprolog
%%CAPTION
 ,caption={\textit{PartitionCHCs}: Translation to CHCs of the \textit{Partition} program. 
 %
}
 ,captionpos=b
 ,label={lst:partitionchc1}
 ,abovecaptionskip=6pt
 ,belowcaptionskip=4pt
] 
all_grt(X,[],B) :- X>=0, B=true.
all_grt(X,[Y|Ys],B) :- X=<Y, X>=0, B=false.
all_grt(X,[Y|Ys],B) :- X>Y, Y>=0, all_grt(X,Ys,B).
 
all_leq(X,[],B) :- X>=0, B=true.
all_leq(X,[Y|Ys],B) :- X>Y, Y>=0, B=false.
all_leq(X,[Y|Ys],B) :- X=<Y, X>=0, all_leq(X,Ys,B).
 
partition(X,[],[],[]).
partition(X,[Y|Ys],[Y|L1s],L2s) :-  X>Y, Y>=0, partition(X,Ys,L1s,L2s).
partition(X,[Y|Ys],L1s,[Y|L2s]) :-  X=<Y, X>=0, partition(X,Ys,L1s,L2s).
\end{lstlisting}

\noindent
The atoms
(i) \texttt{all_grt(X,L,B)}, (ii) \texttt{all_leq(X,L,B)}  and (iii) \texttt{partition(X,L,L1,L2)}  hold in the least model of 
\textit{PartitionCHCs}
iff
the 
expressions (i) \texttt{all_grt(X,L)==B}, (ii) \texttt{all_leq(X,L)==B}  and (iii)~\texttt{partition(X,L)==(L1,L2)}, 
 respectively, 
hold in the functional program \textit{Partition} of Listing~\ref{lst:partition}. 
Contract \ispart is translated into the following two goals \texttt{G1} and \texttt{G2},
whose conjunction is equivalent to the contract: 

\begin{lstlisting}[style=myprolog
%%CAPTION
 ,caption={CHC translation of the \ispart contract.}
 ,captionpos=b
 ,label={lst:partitionchc2}
 ,abovecaptionskip=6pt
 ,belowcaptionskip=4pt
   ,escapechar=*
] 
false :- B=false, partition(X,L,L1,L2), all_grt(X,L1,B). *\hfill*% G1*~~~~*   
false :- B=false, partition(X,L,L1,L2), all_leq(X,L2,B). *\hfill*% G2*~~~~*   
\end{lstlisting}



By a slight abuse of notation we use \texttt{false} to denote both the empty disjunction
in the conclusion of a clause and a boolean value in a constraint.
However, these two uses of \texttt{false} never generate any confusion. 
The use of the constraint \texttt{B=false} allows us to avoid negative literals 
in the body of goals, and hence to stick to Horn format.
The satisfiability of $\textit{PartitionCHCs}\cup\{G\}$, for all $G\in\{\texttt{G1,G2}\}$,
guarantees that \texttt{partition} satisfies the contract \ispart.

Let us consider $\textit{PartitionCHCs}\cup\{\texttt{G2}\}$
{(the satisfiability of 
$\textit{PartitionCHCs}\cup\{\texttt{G1}\}$ can be proved in a similar way)}.
Unfortunately, $\textit{PartitionCHCs}\cup\{\texttt{G2}\}$
cannot be proved satisfiable by state-of-the-art CHC solvers, such as Eldarica or Z3, 
because they do not use any method, such as induction on the list structure, which would be needed
for reasoning on universally quantified list properties (goals, and in general clauses,
have an implicit universal quantification in front).

To overcome this difficulty, we now apply the Elimination Algorithm, which uses the {\em definition}, 
{\em unfolding}, and {\em folding} rules~\cite{EtG96,TaS84},
and from $\textit{PartitionCHCs}\cup\{\texttt{G2}\}$ we derive an equisatisfiable set ${T_{\texttt{G2}}}$
of CHCs where lists do not occur.  
In this way,  we can check the satisfiability of the transformed CHCs  ${T_{\texttt{G2}}}$ 
using a solver over \textit{Bool} and \textit{LIA}, 
without the need for any induction-based method for reasoning on lists.
%
%
%
%
%
We start off by introducing a new predicate \texttt{pl} defined by the following clause
(the variable names are automatically generated by the interactive transformation system MAP~\cite{Re&98}):
%

\begin{lstlisting}[style=myprolog]
1.  pl(A,B) :- partition(B,C,D,E), all_leq(B,E,A).
\end{lstlisting}
and we eliminate list terms from goal \texttt{G2} by folding it using clause~1 as follows:
\begin{lstlisting}[style=myprolog]
F.  false :- B=false, pl(X,B). 
\end{lstlisting}
Now, we look for a recursive definition of predicate \texttt{pl} without occurrences of lists.
By unfolding clause~1
with respect to the \texttt{partition} atom,
we obtain
\begin{lstlisting}[style=myprolog]
2.  pl(A,B) :- all_leq(B,[],A).
3.  pl(A,B) :- B>=0, partition(B,D,E,F), all_leq(B,F,A).
4.  pl(A,B) :- B=<C, B>=0 partition(B,D,E,F), all_leq(B,[C|F],A).
\end{lstlisting}
We proceed by unfolding clauses~2
and~4
with respect to \texttt{all_leq} atoms, 
thereby obtaining

%
\begin{lstlisting}[style=myprolog]
5.  pl(A,B) :- A=true, B>=0.
6.  pl(A,B) :- B=<C, B>=0, partition(B,D,E,F), B>C, A=false.
7.  pl(A,B) :- B=<C, B>=0, partition(B,D,E,F), B=<C, all_leq(B,F,A).
\end{lstlisting}
We remove clause~6
because it contains an unsatisfiable constraint.
Moreover, clause~7 is equal to clause~3, modulo equivalence of constraints,
and thus we remove it.
As a final step, 
we use the definition clause~1 
for folding  clause~3, 
hence deriving
the following final set ${T_{\texttt{G2}}}$ of CHCs:
\begin{lstlisting}[style=myprolog]
5.  pl(A,B) :- A=true, B>=0.
8.  pl(A,B) :- B>=0, pl(A,B).
F.  false :- A=false, pl(A,B).
\end{lstlisting}

The set ${T_{\texttt{G2}}}$ is satisfiable, and Eldarica easily finds that \texttt{pl(A,B)\,:-\,A=true,\,B>=0}  is a {\em model} for  ${T_{\texttt{G2}}}$.
Indeed, by replacing each occurrence of \texttt{pl(A,B)} by \texttt{(A=true,\,B>=0)} in the clauses of ${T_{\texttt{G2}}}$, we derive
clauses that are true in the combined theory of booleans and integers.

%% file: sections/3_specification.tex
\label{spec}

Now, we consider the following program that implements the \textit{Quicksort} algorithm:

\medskip
\begin{lstlisting}[style=myscala
%%CAPTION
 ,caption={Program \textit{Quicksort}.}
 ,captionpos=b
 ,label={lst:quicksort}
 ,abovecaptionskip=7pt
 ,belowcaptionskip=4pt
 ,escapechar=|
]
def quicksort(l: List[Nat]): List[Nat] = {
  l match {
    case Nil() => Nil[Nat]()
    case Cons(x, xs) =>
      val (ys,zs) = partition(x, xs)
      append(quicksort(ys), Cons(x, quicksort(zs)))
  }
} ensuring { res =>  
    forall((a: Nat) => all_grt(a,l) ==> all_grt(a,res)) && |\label{line:qs1}|
    forall((a: Nat) => all_leq(a,l) ==> all_leq(a,res)) &&
    isSorted(0,res) &&
    forall((a: Nat) => count(a,l) == count(a,res))         |\label{line:qs4}|
}

def append(l: List[Nat], ys: List[Nat]): List[Nat] = {
  require( isSorted(0,l) && ( ys == Nil() ||
    ( all_grt(ys.head,l) && all_leq(ys.head,ys.tail) && isSorted(0,ys.tail) )))
  l match {
    case Nil() => ys
    case Cons(x, xs) => Cons(x, append(xs,ys))
  }
} ensuring { res => isSorted(0,res) }
\end{lstlisting}

\noindent
In program \textit{Quicksort}, the function \texttt{partition} is defined as in
Listing~\ref{lst:partition}.
The variable \texttt{res} denotes the return value of a given function.
The \texttt{\color{blue}\bfseries require} and \texttt{\color{blue}\bfseries ensuring} assertions 
specify the
{preconditions and postconditions of} the
contracts for the \texttt{quicksort} and  \texttt{append} functions.
The contract specifications use the functions \texttt{all_grt} and \texttt{all\_leq}
defined in Listing~\ref{lst:partition}, and also the functions
\texttt{count} and \texttt{isSorted} defined below.

\smallskip
\begin{lstlisting}[style=myscala
%%CAPTION
 ,captionpos=b
 ,label={lst:quicksortAuxfun}
 ,abovecaptionskip=3pt
 ,belowcaptionskip=3pt
 ,caption={Auxiliary functions for the \textit{Quicksort} contracts.}
] 
def count(a: Nat, l: List[Nat]): Nat = {
  l match {
    case Nil() => 0
    case Cons(x, xs) => if (x==a) { count(a,xs)+1 } else { count(a,xs) }
  }
}

def isSorted(a: Nat, l: List[Nat]): Boolean = {
  l match {
    case Nil() => true
    case Cons(x,xs) => if (a<=x) isSorted(x,xs) else false
  }
}
\end{lstlisting}
All of the functions used in the contract specifications
have a common recursive pattern, which slightly extends
the {\em catamorphism} pattern defined in functional programming~\cite{MeijerFP91}.
Indeed, the functions considered here admit an extra parameter, and are called
{\em parameterized catamorphisms}.
In particular, the function \texttt{isSorted} is defined by
induction on the list structure by considering the two cases where the input
list \texttt{l} is either \texttt{Nil()} or \texttt{Cons(x,xs)}. By using
the extra parameter \texttt{a} we avoid to split the case \texttt{Cons(x,xs)}
into \texttt{Cons(x,Nil())} and \texttt{Cons(x,Cons(y,xs))},
and we express the sortedness of list \texttt{l} as \texttt{isSorted(0,l)}
(recall that the elements of \texttt{l} are all nonnegative numbers).
The general pattern of parameterized catamorphisms is defined below.

\smallskip
\begin{lstlisting}[style=myscala
%  xleftmargin=0pt
% ,language=scala
%%%KEYWORDS
% ,keywordstyle=\color{blue}\bfseries
% ,alsoletter={<>=-+|&}
%%,morekeywords={ensuring,require,==>,||,&&,=>}
% ,morekeywords={ensuring,require}
% ,otherkeywords={}
% ,deletekeywords={true,false}
%%LINE NUMBERS
 ,numbers=left
 ,numberstyle=\sffamily\tiny
 ,numbersep=15pt
 ,numberblanklines=false
%,firstnumber=100
 ,firstnumber=last
 ,numbers=none
%%CAPTION
 ,captionpos=b
 ,abovecaptionskip=2pt
 ,belowcaptionskip=4pt
 ,label={lst:generalcata}
 ,caption={General form of parameterized catamorphism on \texttt{List[A]}.}
] 
def pCata(p:A, l:List[A]): B = {
  match l {
    case Nil() => c
    case Cons(x,xs) => g(p,x,pCata(h(p,x),xs))
  }
}
\end{lstlisting}
In Listing~\ref{lst:generalcata}, 
(i)~\texttt{A} is any type and \texttt{B} is the type of the integer or boolean values,
(ii)~\texttt{c} is a constant of type \texttt{B},
(iii)~\texttt{g} is a {total}, \texttt{B}-valued function, and
(iv)~\texttt{h} is a {total}, \texttt{A}-valued function.
Thus, also \texttt{pCata} is a total, \texttt{B}-valued function.

\newcommand{\tvar}[1] { \mathtt{\bar{#1}} }

The task of verifying the contract for a function \texttt{f} consists in 
proving the validity of a universally quantified implication of the form:
\begin{lstlisting}[style=myscala,numbers=none]
  $\forall\tvar{x}$. pre($\tvar{x}$) ==> post($\tvar{x}$,f($\tvar{x}$))
\end{lstlisting}
where:
(i) $\tvar{x}$ is a tuple of variables (a subset of the function inputs), and
(ii) \texttt{pre($\tvar{x}$)} and \texttt{post($\tvar{x}$,f($\tvar{x}$))} are 
the precondition and postcondition, respectively, specified by the \texttt{\color{blue}\bfseries require} 
and \texttt{\color{blue}\bfseries ensuring} assertions
using parameterized catamorphisms.

The \texttt{pre($\tvar{x}$)} assertion for \texttt{quicksort} is absent.
Thus, verifying the contract of \texttt{quicksort} consists in 
verifying the validity of~
$\forall\texttt{l.\,true ==> post(l,quicksort(l))}$,
where \texttt{post(l,quicksort(l))} is the conjunction of the
following assertions: 

\begin{enumerate}[topsep=0pt,partopsep=0pt,parsep=0pt,itemsep=-10pt]
\item[\texttt{1.}]\label{f1}
\begin{lstlisting}[style=myscala,numbers=none]
$\forall$a. all_grt(a,l) ==> all_grt(a,quicksort(l)))
\end{lstlisting}

\item[\texttt{2.}]\label{f2}
\begin{lstlisting}[style=myscala,numbers=none]
$\forall$a. all_leq(a,l) ==> all_leq(a,quicksort(l)))
\end{lstlisting}

\item[\texttt{3.}]\label{f3}
\begin{lstlisting}[style=myscala,numbers=none]
isSorted(0,quicksort(l))
\end{lstlisting}

\item[\texttt{4.}]\label{f4}
\begin{lstlisting}[style=myscala,numbers=none]
$\forall$a. count(a,l) == count(a,quicksort(l)))
\end{lstlisting}
\end{enumerate}

\noindent
Assertions 1 and 2 state that \texttt{quicksort} preserves
the postcondition of the function \texttt{partition}.
Assertion 3 expresses the sortedness property.
Assertion 4 states that the multiset of natural numbers in the input list 
\texttt{l} is the same as the multiset of the elements in 
\texttt{quicksort(l)}.

For the function \texttt{append}, the precondition \texttt{pre(l,ys)},
where \texttt{l} and \texttt{ys} are the input lists,
 is defined as follows:

\begin{lstlisting}[style=myscala,numbers=none,lineskip=1pt]
      isSorted(0,l) && ( ys == Nil() ||
        ( all_grt(ys.head,l) && all_leq(ys.head,ys.tail) && isSorted(0,ys.tail) ) )
\end{lstlisting}

\noindent
The assertion states that (i) \texttt{l} is sorted, and either (ii) \texttt{ys}
is the empty list or
(iii.1) the head of \texttt{ys} is greater than every element of \texttt{l},
(iii.2) the head of \texttt{ys} is less than or equal to every element occurring in its tail, and
(iii.3) the tail of \texttt{ys} is sorted.
The postcondition of the function \texttt{append} states that its output is a sorted list.

\smallskip
The \textit{Quicksort} program (Listing~\ref{lst:quicksort}) and
the auxiliary functions (Listing~\ref{lst:quicksortAuxfun})
are translated to the set \textit{QuicksortCHCs} of clauses 
in Listing~\ref{lst:chcqs} below.

\begin{lstlisting}[style=myprolog
%%CAPTION
 ,captionpos=b
 ,abovecaptionskip=6pt
 ,belowcaptionskip=4pt
 ,label={lst:chcqs}
 ,caption={\textit{QuicksortCHCs}: CHC translation of \textit{Quicksort} and its auxiliary functions.}
]
quicksort([],[]).
quicksort([X|Xs],Ys) :- X>=0,
  partition(X,Xs,Littles,Bigs),
  quicksort(Littles,Ls), quicksort(Bigs,Bs),
  append(Ls,[X|Bs],Ys).

append([],Xs,Xs).
append([X|Xs],Ys,[X|Zs]) :- X>=0, append(Xs,Ys,Zs).

count(X,[],N) :- X>=0, N=0.
count(X,[Y|Ys],N) :- X>=0, X=Y, N=M+1, count(X,Ys,M).
count(X,[Y|Ys],N) :- X>=0, Y>=0, X=\=Y, N=M, count(X,Ys,M).

isSorted(A,[],B) :- A>=0, B=true.
isSorted(A,[X|Xs],B) :- X>=0, A>X, B=false.
isSorted(A,[X|Xs],B) :- A>=0, A=<X, isSorted(X,Xs,B).
\end{lstlisting}

The contracts are translated to CHC goals as follows.

\begin{lstlisting}[style=myprolog
%%CAPTION
 ,captionpos=b
 ,abovecaptionskip=6pt
 ,belowcaptionskip=4pt
 ,label={lst:chccontracts}
 ,caption={CHC translation of the \textit{Quicksort} contracts.}
 ,escapechar=*
 ,lineskip=0.1mm
]
% quicksort contract
false :- B1=true, B2=false, all_grt(A,B,B1), quicksort(B,C), all_grt(A,C,B2). *\hfill*% G3*~~~~*    
false :- B1=true, B2=false, all_leq(A,B,B1), quicksort(B,C), all_leq(A,C,B2). *\hfill*% G4*~~~~*
false :- B1=false, quicksort(L,S), isSorted(0,S,B1). *\hfill*% G5*~~~~*
false :- N1=/=N2, count(X,L,N1), quicksort(L,S), count(X,S,N2). *\hfill*% G6*~~~~*

% append  contract
false :- B1=true, B2=true, B3=true, B4=true, B5=false,*\hfill*% G7*~~~~*
  all_grt(X,Xs,B1), isSorted(0,Xs,B2),
  all_leq(X,Ys,B3), isSorted(0,Ys,B4),
  append(Xs,[X|Ys],Zs), isSorted(0,Zs,B5).
\end{lstlisting}
Similarly to the translation of the contract for the \textit{Partition} program,
the use of boolean constraints avoids the introduction of negative literals.

Now, to prove the correctness of \textit{Quicksort}
with respect to its contracts, it remains to show that $\textit{QuicksortCHCs}\cup \{G\}$ is satisfiable
for all goals $G\in \{ \texttt{G3,G4,G5,G6,G7} \}$. 
Unfortunately, these satisfiability problems cannot be directly solved by Eldarica or Z3.


%
%

%% file: sections/4_transformation.tex
\label{listrem}

Similarly to the \texttt{partition} example presented in Section~\ref{sec:partition},
the proof of satisfiability of the set of clauses $\textit{QuicksortCHCs}\cup \{G\}$,
where $G$ is a goal among \texttt{G3,G4,G5,G6,G7},
proceeds in two steps.
First, we transform $\textit{QuicksortCHCs}\cup \{G\}$ by using the fold/unfold rules,
and derive a new set $T_G$ such that:
(i) $T_G$ is a set of CHCs over \textit{LIA} and \textit{Bool}, without any list argument, and
(ii) if $T_G$ is satisfiable, then $\textit{QuicksortCHCs}\cup \{G\}$ is satisfiable.
Then, we check the satisfiability of $T_G$ by using a CHC solver  over \textit{LIA} and \textit{Bool}.

The main difference with respect to the \texttt{partition} example is
that we also use as lemmas the  properties that we have
already proved in previous applications of our method.
For instance, having proved that $\textit{PartitionCHCs}\cup \{\texttt{G2}\}$
is satisfiable (see Section~\ref{sec:partition}), during subsequent transformations we can use the property 

\begin{lstlisting}[style=myprolog]
$\forall \texttt{X,L,L1,L2}$. partition(X,L,L1,L2) ==> all_leq(X,L2,true)
\end{lstlisting}
and add (instances of) \texttt{all\_leq(X,L2,true)} to the body of a clause 
where \texttt{partition(X,L,L1,L2)} occurs.

The general form of the transformation strategy that we apply to eliminate list terms
is an extension of the {Elimination Algorithm}~\cite{De&18a}.
The strategy is parametric with respect to specific 
$\mathit{Define\mbox{-}Fold}$,
$\mathit{Unfold}$, and $\mathit{Replace}_{cata}$ functions.

\noindent \hrulefill\nopagebreak

\noindent {\bf List Removal}~$\mathcal R_{cata}$.\\
{\em Input}: A set $\mathit{Cls} \cup \{G\}$, where $\mathit{Cls}$ is a set of non-goal clauses and $G$ is a {goal},
and a set $\mathit{Props}$ of properties in the form of implications \texttt{B1 ==> B2};
\\
{\em Output}: A set $T_G$ of clauses over \textit{LIA} and \textit{Bool} such that if $T_G$ is satisfiable, then $\mathit{Cls} \cup \{G\}$  is satisfiable.

\vspace*{-2mm}
\noindent \rule{2.0cm}{0.2mm}

\noindent $\mathit{Defs}:=\emptyset$;~~
\noindent $\mathit{InCls}:=\{G\}$;~~
\noindent $\mathit{T_G}:=\emptyset;$

\noindent
{\bf while} $\mathit{InCls}\!\neq\!\emptyset$ {\bf do}

$(\mathit{NewDefs},\mathit{FldCls}) := \mathit{Define\mbox{-}Fold}(\mathit{Defs},\mathit{InCls});$

$\mathit{UnfCls} := \mathit{Unfold}(\mathit{NewDefs},\mathit{Cls});$

$\mathit{RCls} := \mathit{Replace}_{cata}(\mathit{UnfCls},\mathit{Props});$

$\mathit{Defs}:=\mathit{Defs}\cup\mathit{NewDefs};$~~
$\mathit{InCls}:=\mathit{RCls};$~~
$\mathit{T_G}:=\mathit{T_G}\cup\mathit{FldCls};$

\vspace*{-2mm} 
\noindent \hrulefill

\smallskip

In $\mathcal R_{cata}$, the set \textit{Defs} of clauses
stores the new definitions 
introduced during the application of the transformation strategy. 
The set $\mathit{InCls}$ is the set of clauses to be transformed.
$T_G$ is the set of transformed clauses.
$\mathit{NewDefs}$ and $\mathit{FldCls}$ are the sets of clauses derived by
applying the definition and folding rules, respectively using
the function $\mathit{Define\mbox{-}Fold}$.
$\mathit{UnfCls}$ is the set of clauses derived by applying the unfolding rule
using the function $\mathit{Unfold}$.
$\mathit{RCls}$ is the set of clauses derived by 
applying the function $\mathit{Replace}_{cata}$, which
uses properties stored in $\mathit{Props}$ 
corresponding to goals whose satisfiability has been proved in previous steps.

Let us explain the list removal strategy in action for the transformation of 
$\textit{QuicksortCHCs}\cup \{\texttt{G5}\}$, where we also use
the properties \textit{Props} corresponding to goals \texttt{G1,G2,G3,G4}.
The properties corresponding to goals \texttt{G6} and  \texttt{G7} are not needed
for~\texttt{G5}.

\smallskip
\noindent
$\mathit{Define\mbox{-}Fold}$. \
$\mathcal R_{cata}$ starts off by introducing the following new predicate:
\begin{lstlisting}[style=myprolog]
1.  qss(A) :- quicksort(B,C), isSorted(0,C,A).
\end{lstlisting}
and folding the goal \texttt{G5} as follows:
\begin{lstlisting}[style=myprolog]
F5.  false :- A=false, qss(A).
\end{lstlisting}
$\mathit{Unfold}$. \ By unfolding clause 1 with respect to the \texttt{quicksort} and the \texttt{isSorted} atoms, we get:
\begin{lstlisting}[style=myprolog]
2. qss(true).
3. qss(A) :- B>=0,
	partition(B,C,D,E), quicksort(D,F), quicksort(E,G), 
	append(F,[B|G],H), isSorted(0,H,A).
\end{lstlisting}
$\textit{Replace}_{cata}$. \ Now, we apply the properties corresponding 
to goals \texttt{G1} and \texttt{G2}, which translate the
postcondition of the \texttt{partition} function (see Section~\ref{sec:partition}), and we add the two atoms
\texttt{all\_grt(B,D,true)} and \texttt{all\_leq(B,E,true)} to the body of clause 3:
\begin{lstlisting}[style=myprolog]
4. qss(A) :- B>=0, 
	partition(B,C,D,E), all_grt(B,D,true), all_leq(B,E,true), 
	quicksort(D,F), quicksort(E,G), append(F,[B|G],H), isSorted(0,H,A).
\end{lstlisting}
Next, by using \texttt{G3} and \texttt{G4}, 
{we add the atoms
\texttt{all\_grt(B,F,true)} and \texttt{all\_leq(B,G,true)} to the body of clause 4 and}
we derive:
\begin{lstlisting}[style=myprolog]
5. qss(A) :- B>=0, 
	partition(B,C,D,E), all_grt(B,D,true), all_leq(B,E,true), 
	quicksort(D,F), all_grt(B,F,true), 
	quicksort(E,G), all_leq(B,G,true), 
	append(F,[B|G],H), isSorted(0,H,A).
\end{lstlisting}
Now, in order to fold the two \texttt{quicksort} atoms using clause 1,
we add two instances of the parameterized catamorphism \texttt{isSorted},
where the output boolean value is an unbound variable, and hence
implicitly existentially quantified.
This step is correct because, by the totality of the \texttt{isSorted}
function, the following property holds: \texttt{$\forall$L:List[Nat]$\,\exists$B:Boolean.\,isSorted(0,L,B)}.

%
Hence, we get:
\begin{lstlisting}[style=myprolog]
6. qss(A) :- B>=0, 
	partition(B,C,D,E), all_grt(B,D,true), all_leq(B,E,true), 
	quicksort(D,F), isSorted(0,F,B1), all_grt(B,F,true), 
	quicksort(E,G), isSorted(0,G,B2),  all_leq(B,G,true), 
	append(F,[B|G],H), isSorted(0,H,A).
\end{lstlisting}
Note that we cannot use the property corresponding to goal \texttt{G7}
because \texttt{B1} and \texttt{B2} are unbound variables, while 
\texttt{G7} requires them to be bound to \texttt{true}.

Now, we perform a second iteration of the List Removal strategy.

\smallskip
\noindent
$\mathit{Define\mbox{-}Fold}$. \
We fold twice clause 6 using clause 1, and we get:
\begin{lstlisting}[style=myprolog]
7. qss(A) :- B>=0, 
	partition(B,C,D,E), all_grt(B,D,true), all_leq(B,E,true), 
	qss(B1), isSorted(0,F,B1), all_grt(B,F,true), 
	qss(B2), isSorted(0,G,B2), all_leq(B,G,true), 
	append(F,[B|G],H), isSorted(0,H,A).
\end{lstlisting}
By this folding step, we do not remove the \texttt{isSorted} atoms,
which share the lists \texttt{F} and \texttt{G} with
the \texttt{append} atom.
In contrast, we  remove the conjunction 
\texttt{partition(B,C,D,E),all_grt(B,D,true),} \texttt{all_leq(B,E,true)},
which, by the totality of \texttt{partition(B,C,D,E)} and by 
the {properties corresponding to}
goals
\texttt{G1} and \texttt{G2}, is always true:
\begin{lstlisting}[style=myprolog]
8. qss(A) :- B>=0, 
	qss(B1), isSorted(0,F,B1), all_grt(B,F,true), 
	qss(B2), isSorted(0,G,B2), all_leq(B,G,true), 
	append(F,[B|G],H), isSorted(0,H,A).
\end{lstlisting}
Then, we introduce the following new definition:
\begin{lstlisting}[style=myprolog]
9. a(B,X,Y,Z,T,U,B1,B2,A) :- 
	isSorted(X,F,B1), all_grt(B,F,T), 
	isSorted(Y,G,B2), all_leq(B,G,U), 
	append(F,[B|G],H), isSorted(Z,H,A).
\end{lstlisting}
which we use for folding clause 8, hence deriving: 
\begin{lstlisting}[style=myprolog]
10. qss(A) :- B>=0, qss(B1), qss(B2), a(B,0,0,0,true,true,B1,B2,A).
\end{lstlisting}
Now, predicate \texttt{qss} is defined by clauses
2 and 10, which have no lists.
However, predicate \texttt{a}, occurring in the body of clause 10,
is defined by clause 9, whose body has some occurrences of list terms.
Thus, the List Removal strategy continues by transforming clause 9
and, after a few iterations, produces a set of clauses without lists.
The final result of this transformation is a set $T_{\texttt{G5}}$ including goal
\texttt{F5}, clauses 2 and 10, and the clauses for predicate
\texttt{a} (and some extra predicates introduced in subsequent iterations)
reported in the Appendix.
$T_{\texttt{G5}}$ is a set of Horn clauses  with constraints in \textit{LIA} and~\textit{Bool} {only}.

The CHC solver Eldarica is able to prove the satisfiability of $T_{\texttt{G5}}$, 
and hence also the initial set of clauses $\textit{QuicksortCHCs}\cup \{\texttt{G5}\}$
is satisfiable. 
Similarly, by applying again the List Removal strategy and then
proving satisfiability by Eldarica over \textit{LIA} and \textit{Bool}, 
we are able to verify all contracts of the \textit{Quicksort}
program.

We have also attempted to verify the same contracts by using the
{\sc Stainless} system~\cite{HamzaVK19}, a verifier for the Scala language.
{\sc Stainless} is able to verify the contracts of the functions \texttt{partition} and
\texttt{append}, but not the one of \texttt{quicksort}.

%% file: sections/5_relConcl.tex
\label{relwork}
The \textit{Quicksort} algorithm is a brilliant invention by Tony Hoare, presented in
his famous 1961 paper~\cite{Hoare61}.
A formal proof of partial correctness, using the axiomatic approach~\cite{Hoa69},
was presented by Hoare himself, in a joint paper with M. Foley~\cite{FoleyH71}.
Since then, many hand-made proofs have been worked out, for several 
variants (both recursive and iterative) of the algorithm (see, for instance,
the book by Apt et al.~\cite{Ap&09}).
Also semi-automated proofs have been presented, using {\em program verifiers}
that implement Hoare logic, such as
{\sc Dafny}~\cite{CertezeanuDELSW16,Lei13} and {\sc Stainless}~\cite{HamzaVK19}.
However, the success of program verifiers is very much dependent on
the assertions provided by the programmer.
In particular, we have checked that  {\sc Stainless} is able to verify
the contracts of a program implementing a variant of  \textit{Quicksort}~\footnote{See the verification benchmarks
at~~\url{https://github.com/epfl-lara/stainless/}.},
but it could not verify the version presented in Section~\ref{spec}
of this paper.
 
Also our proof  depends critically on the contract specifications, because
we first prove and then use them as lemmas during the transformation phase. 
For instance, a crucial role is played by the postcondition of the 
\texttt{partition} function, that is, contract \ispart of Section~\ref{sec:partition}:

\smallskip
\noindent
~~$\forall$\texttts{x,l,l1,l2. partition(x,l)==(l1,l2) 
	==> all_grt(x,l1)  \&\&   all_leq(x,l2) }

\smallskip
\noindent
stating that the output of \texttt{partition} is a pair of lists \texttt{(l1,l2)}
such that  the pivot \texttt{x} is greater than all elements in \texttt{l1},
and smaller or equal than all elements in \texttt{l2}.
Without introducing the two predicates \texttt{all\_grt} and  \texttt{all\_leq},
and then proving that they are preserved by applications of the 
\texttt{quicksort} function,
our transformation would not go through.

Another interesting point is that in all contract specifications we 
use predicates defined by a simple induction scheme, 
which we have called parameterized catamorphisms. 
This form helps introducing suitable new predicates (the famous {\em eureka definitions}
in Burstall and Darlington's approach~\cite{BuD77}).
Indeed, all predicates introduced by the definition rule in our transformations (including the
ones not shown in the paper) are defined as a conjunction
of an atom, representing a call to a program function, and one or more atoms representing 
parameterized catamorphism.
We argue that, by exploiting properties of parameterized catamorphisms, one can develop a fully automatic 
version of the transformation
strategy~$\mathcal R_{cata}$ that always succeeds in eliminating lists and, more in general,
inductively defined data structures, from large classes of CHCs. We leave this task for future research.

Catamorphisms (on trees) were used in the context of {\em Satisfiability Modulo Theories}, to define satisfiability 
algorithms that terminate for suitable classes of formulas~\cite{PhW16,SuterDK10}.
A special form of integer-valued catamorphisms, such as {\em list length}, {\em term-size}, and 
in general, the so-called {\em type-based norms}, are used by techniques for
proving termination of logic programs~\cite{BruynoogheCGGV07}.
Our definition of parameterized catamorphism slightly extends the one
of list catamorphism usually given in the context of functional programming~\cite{MeijerFP91}.
Our definition allows an extra parameter, which makes the inductive scheme a little more flexible.

A more challenging problem is to discover pre/postconditions defined by catamorphisms
which are not provided by the programmer. For instance, suppose that for the \textit{Quicksort} program
the programmer only specifies the contract for the main function \texttt{quicksort}
using the functions \texttt{isSorted} and \texttt{count}. Then, an automated 
verifier (or transformer) should be able to discover suitable pre/postconditions such as
the ones we have provided in terms of predicates \texttt{all\_grt} and  \texttt{all\_leq}.
This problem is related to the discovery of suitable lemmata
during automated theorem proving~\cite{Bun01} and program transformation~\cite{De&20a}, which is well-known
to be very hard.
However, we argue that, restricting the search for those lemmata among (parameterized)
catamorphisms of suitable form, could be a fruitful heuristic.

\section*{Acknowledgment}

We shared most of our scientific careers, starting from their beginnings, with Alberto Pettorossi,
and the influence of his approach to Science on our own way of doing research has been enormous.
First of all, the topics he contributed to explore starting from the 1970s, such as program transformation,
program verification and, in general, the use of logic and formal methods in computing, are still
central in our current research, as witnessed by the present paper.
But, much more than that, Alberto's everyday example taught us the commitment and honesty in doing research,
looking at the substance and not at the surface, without, however, neglecting form and beauty.
Indeed, Alberto's classical studies at secondary school, Latin and Greek ancient languages, as well as
Italian literature classics, had a big impact on his way of writing papers.
``We must love our readers" is one of his recurrent sentences!
Finally, we want to say that Alberto's teachings go far beyond the scientific side: 
through his continuous emotional support of young and weak people,
he has always shown the joy of committing one's life to something valuable.   

Thanks Alberto, our teacher, colleague and friend.

\smallskip

We would also like to thank Laurent Fribourg, Alexei Lisitsa, and Andrei Nemytykh, for
organizing this workshop dedicated to Alberto and inviting us to write this paper.
This work has been partially supported by INdAM-GNCS.

%% file: sections/7_appendix.tex
\label{appendix}

We list below the final set $T_{\texttt{G5}}$ of clauses derived from
$\textit{QuicksortCHCs}\cup \{\texttt{G5}\}$.
Clauses 11--28 have been derived automatically from clause 9, by using the implementation of
the Elimination Algorithm on the VeriMAP~\footnote{The tool is available at  
\url{https://fmlab.unich.it/iclp2018/}.} system~\cite{De&14b}.
Both Eldarica and Z3 are able to prove the satisfiability of this set of clauses.

\begin{lstlisting}[style=myprolog]
F5. false :- A=false, qss(A).
2.   qss(A) :- A=true.
10.  qss(A) :- B>=0, qss(B1), qss(B2), a(B,0,0,0,true,true,B1,B2,A).
11.  a(A,B,C,D,E,F,G,H,I) :- A=J, B=0, C=0, D=0, E=true, F=true, G=K, H=L, I=M, N=0,
		O=0, P=0, Q=J, R=true, S=J, T=true, J>=0, new2(J,T,Q,S,R,P,K,O,L,N,M).
12.  new2(A,B,C,D,E,F,G,H,I,J,K) :- A=L, B=true, C=L, D=L, E=true, F=0, G=true, H=0, 
		J=0, K=M, L>=0, new3(L,M,D,E,H,I).
13.  new2(A,B,C,D,E,F,G,H,I,J,K) :- A=L, B=true, C=L, D=L, E=true, F=0, G=M, H=0, 
		J=0, K=N, O=true, P=Q, Q-L=< -1, Q>=0, new6(C,L,O,P,M,Q,N,D,E,H,I).
14.  new3(A,B,C,D,E,F) :- A=C, B=true, D=true, E=0, F=true, C>=0.
15.  new3(A,B,C,D,E,F) :- A=G, B=H, C=G, D=true, E=0, F=H, I=true, G>=0, J-G>=0, 
		new10(G,I,J,H).
16.  new6(A,B,C,D,E,F,G,H,I,J,K) :- A=L, B=L, C=true, D=F, E=true, G=M, H=L, I=true, 
		J=0, F>=0, L-F>=0, new7(L,M,H,I,J,K).
17.  new6(A,B,C,D,E,F,G,H,I,J,K) :- A=L, B=L, C=true, D=F, E=false, G=false, H=L, 
		I=true,	J=0, M=true, F>=1, L-F>=0, new9(A,L,M,H,I,J,K).
18.  new6(A,B,C,D,E,F,G,H,I,J,K) :- A=L, B=L, C=true, D=F, E=M, G=N, H=L, I=true, 
		J=0, O=true, P=Q, Q-L=< -1, F>=0, Q-F>=0, new6(A,L,O,P,M,Q,N,H,I,J,K).
19.  new7(A,B,C,D,E,F) :- A=C, B=true, D=true, E=0, F=true, C>=0.
20.  new7(A,B,C,D,E,F) :- A=G, B=H, C=G, D=true, E=0, F=H, I=true, G>=0, J-G>=0, 
		new10(G,I,J,H).
21.  new9(A,B,C,D,E,F,G) :- A=D, B=D, C=true, E=true, F=0, G=true, D>=1.
22.  new9(A,B,C,D,E,F,G) :- A=H, B=H, C=true, D=H, E=true, F=0, G=I, J=true, H>=1,
        K>=H, new10(H,J,K,I).
23.  new9(A,B,C,D,E,F,G) :- A=H, B=H, C=true, D=H, E=true, F=0, I=true, H>=1, 
		new9(A,H,I,D,E,F,G).
24.  new10(A,B,C,D) :- B=true, D=true, A>=0, C-A>=0.
25.  new10(A,B,C,D) :- A=E, B=true, D=false, F=true, E-C=< -1, E>=0, new11(E,F).
26.  new10(A,B,C,D) :- A=E, B=true, D=F, G=true, E-C=<0, E>=0, H>=C, new10(E,G,H,F).
27.  new11(A,B) :- B=true, A>=0.
28.  new11(A,B) :- A=C, B=true, D=true, C>=0, new11(C,D).
\end{lstlisting}
